\pdfoutput=1
\RequirePackage{ifpdf}
\ifpdf % We are running pdfTeX in pdf mode
\documentclass[pdftex]{sigma}
\else
\documentclass{sigma}
\fi
\usepackage[matrix,arrow]{xy}

\numberwithin{equation}{section}

\begin{document}

\allowdisplaybreaks

\renewcommand{\thefootnote}{$\star$}

\renewcommand{\PaperNumber}{021}

\FirstPageHeading

\ShortArticleName{Lagrange Anchor and Characteristic Symmetries of
Free Massless Fields}

\ArticleName{Lagrange Anchor and Characteristic Symmetries\\ of
Free Massless Fields\footnote{This
paper is a contribution to the Special Issue ``Symmetries of Dif\/ferential Equations: Frames, Invariants and Applications''. The full collection is available at \href{http://www.emis.de/journals/SIGMA/SDE2012.html}{http://www.emis.de/journals/SIGMA/SDE2012.html}}}

\Author{Dmitry S.~KAPARULIN, Simon L.~LYAKHOVICH and Alexey A.~SHARAPOV}

\AuthorNameForHeading{D.S.~Kaparilin, S.L.~Lyakhovich and A.A.~Sharapov}

\Address{Department of Quantum Field Theory, Tomsk State University,\\ 36 Lenin Ave., Tomsk 634050, Russia}

\Email{\href{mailto:dsc@phys.tsu.ru}{dsc@phys.tsu.ru},
\href{mailto:sll@phys.tsu.ru}{sll@phys.tsu.ru},
\href{mailto:sharapov@phys.tsu.ru}{sharapov@phys.tsu.ru}}

\ArticleDates{Received December 28, 2011, in f\/inal form April 09, 2012; Published online April 12, 2012}

\Abstract{A Poincar\'e  covariant Lagrange anchor is found for the
non-Lagrangian relativistic wave equations of Bargmann and Wigner
describing free massless f\/ields of spin $s > 1/2$ in
four-dimensional Minkowski space. By making use of this Lagrange
anchor, we assign a~symmetry to each conservation law and perform
the path-integral quantization of the theory.}


\Classification{70S10; 81T70}

\renewcommand{\thefootnote}{\arabic{footnote}}
\setcounter{footnote}{0}

\section{Introduction}
The notions of symmetry and conservation law are of paramount
importance for classical and quantum f\/ield theory. For Lagrangian
theories both the notions are tightly connected to each other due to
Noether's f\/irst theorem. Beyond the scope of Lagrangian dynamics,
this connection has remained unclear, though many particular results
and generalizations are known (see~\cite{K-S} for a~review). In our
recent works~\cite{KLS2,KLS3} a general method has been proposed for
connecting symmetries and conservation laws in not necessarily
Lagrangian f\/ield theories. The key ingredient of the method is the
notion of a \textit{Lagrange anchor} introduced earlier~\cite{KazLS}
in the context of quantization of (non-)Lagrangian dynamics.
Geometrically,  the Lagrange anchor def\/ines a~map from the vector
bundle dual to the bundle of equations of motion to the tangent
bundle of the conf\/iguration space of f\/ields such that certain
compatibility conditions are satisf\/ied. In Lagrangian theories, the
two bundles coincide and one can take the identity map as  a
(canonical) Lagrange anchor. For non-Lagrangian f\/ield equations,
these two bundles may be dif\/ferent and it is a problem by itself to
f\/ind at least one nontrivial Lagrange anchor, let alone the
classif\/ication of all admissible Lagrange anchors.

In this paper,  the general concept of Lagrange anchor is
exemplif\/ied by the Bargmann--Wigner equations for free massless
f\/ields of spin $s\geq 1/2$ in the four-dimensional Minkowski space
\cite{PR}. The choice of the example is not accidental. First of
all, it has long been known that the model admits inf\/inite sets of
rigid symmetries and conservation laws. These have been a~subject of
intensive studies by many authors during decades, see e.g.~\cite{AP0,Fairlie,FN,Kibble,KVZ,Lipkin,Morgan,GSV} and references
therein. However, a complete classif\/ication has been obtained only
recently, f\/irst for the conservation laws~\cite{AP1} and then for
the symmetries~\cite{AP2}. As the f\/ield equations are non-Lagrangian
for~$s> 1/2$, there is no immediate Noether's correspondence between
symmetries and conservation laws. The rich structure of symmetries
and conservation laws in the absence of a Lagrangian formulation
makes this theory an ideal testing area for the concept of Lagrange
anchor.

The main result of the paper is a Poincar\'e invariant Lagrange
anchor for the Bargmann--Wigner equations. It is well known that each
conserved current def\/ines (and is def\/ined by) some characteristic
\cite{BBH, KLS3,Olver}. We have proved earlier~\cite{KLS2} that any
Lagrange anchor maps characteristics to symmetries. For Lagrangian
theories endowed with the canonical Lagrange anchor this map is
actually a bijection in accordance with Noether's f\/irst theorem.
Generally the Lagrange anchor map is neither injective nor
surjective. The symmetries that do originate (by the Lagrange anchor
map) from characteristics  are called \textit{characteristic
symmetries}. In the considered model of the free massless f\/ields,
loosely, the characteristic symmetries cover a ``half'' of
equation's symmetries. They also form a subalgebra in the Lie
algebra of all the symmetries. To the best of our knowledge the
existence of such an inf\/inite dimensional subalgebra in the full Lie
algebra of symmetries has not been noticed before for these well
studied equations. Furthermore, the pull back of the Lie bracket on
characteristic symmetries with respect to the anchor map gives rise
to a Lie  bracket on the space of conservation laws such that the
anchor map appears to be a Lie algebra isomorphism. This bracket
generalizes the Dickey bracket~\cite{D} of the conserved currents in
Lagrangian theory.

For the sake of completeness we also present the quantum probability
amplitude on the space of free massless f\/ields which is determined
by the proposed Lagrange anchor, and which implements the
path-integral quantization of the model.

\section{Equations, symmetries, and characteristics}

\subsection{Field equations}
We consider the free massless f\/ields of
spin $s\geq 1/2$ subject to the relativistic wave-equations
\begin{equation}\label{FEquations}
T^{\dot\alpha}_{\alpha_1\dots\alpha_{2s-1}}:=\partial^{\alpha\dot{\alpha}}\varphi_{\alpha\alpha_1\ldots\alpha_{2s-1}}=0,
\end{equation}
$\varphi_{\alpha_1\ldots\alpha_{2s}}(x)$ being a symmetric,
complex-valued spin-tensor on four-dimensional Minkowski spa\-ce~$\mathbb{R}^{3,1}$. Hereafter we use the standard notation and
conventions of the two-component spinor forma\-lism~\cite{PR}. In
particular,
$\partial^{\alpha\dot{\alpha}}=(\sigma^\mu)^{\alpha\dot{\alpha}}\partial/\partial
x^{\mu}$, $\mu=0,1,2,3$, $\alpha,\dot\alpha=1,2$, and we raise and
lower the spinor indices using the spinor metrics
$\varepsilon_{\alpha\beta}$, $\varepsilon_{\dot\alpha\dot\beta}$ and
their inverse $\varepsilon^{\alpha\beta}$,
$\varepsilon^{\dot\alpha\dot\beta}$. According to the
spin-statistics theorem~\cite{SW}, the f\/ields of integer spin are
considered to be bosonic and the  f\/ields of half-integer spin are
treated as fermionic\footnote{Notice that in~\cite{AP1,AP2}
all the f\/ields $\varphi_{\alpha_1\ldots\alpha_{2s}}$ are treated as
bosonic ones regardless the value of spin.}. Statistics are of no
consequence as long as linear in f\/ields expressions are considered
(such as equations of motion or symmetry transformations); they,
however, become crucial when dealing with nonlinear expressions like
quadratic conserved currents.

For $s\geq 1$ the f\/ield equations (\ref{FEquations}) satisfy the
Noether identities
\begin{equation}\label{NIdentities}
\partial_{\dot\alpha}^{\alpha_1}T^{\dot\alpha}_{\alpha_1\dots\alpha_{2s-1}}\equiv 0  ,
\end{equation}
though there are no gauge symmetries. This indicates that the
equations under consideration are non-Lagrangian save for $s=1/2$,
as in Lagrangian dynamics there is a  one-to-one correspondence
between the gauge symmetries and Noether identities (Noether's
second theorem~\cite{K-S}).

In what follows we will use some terminology of the geometry of jet
spaces, though not systematically.  By the $p$th jet of the f\/ield
$\varphi$ we mean the following collection of space-time functions:
\[
j^p\varphi=\left\{ x^\mu,
\varphi_{\alpha_1\dots\alpha_{2s}}(x),
\partial_{\mu_1}\varphi_{\alpha_1\dots\alpha_{2s}}(x), \ldots,
\partial_{\mu_1}\cdots\partial_{\mu_{p}}\varphi_{\alpha_1\dots\alpha_{2s}}(x)\right\}  .
\]
We say that the space-time function $F$ depends on the $p$th jet of
the f\/ield $\varphi$ if it is a smooth function of the elements of
$j^p\varphi$ considered as independent variables, i.e.,
\[
F=F(x^\mu , \varphi_{\alpha_1\dots\alpha_{2s}}(x),
\partial_{\mu_1}\varphi_{\alpha_1\dots\alpha_{2s}}(x), \ldots,
\partial_{\mu_1}\cdots\partial_{\mu_{p}}\varphi_{\alpha_1\dots\alpha_{2s}}(x)).
\]
We will also refer to $F$ as a \textit{local function of fields}.

\subsection{Symmetries} A variational vector f\/ield
\begin{equation}\label{RSymmetries}
    Z=\int d^4x\left(
    Z_{\alpha_1\dots\alpha_{2s}}\frac{\delta}{\delta
    \varphi_{\alpha_1\dots\alpha_{2s}}(x)}+ \mbox{c.c.}\right)
\end{equation}
is called a \textit{symmetry of order} $p$ if its components
$Z_{\alpha_1\dots\alpha_{2s}}$ depend on the $p$th jet of the f\/ield
$\varphi$ and the following condition is satisf\/ied:
\begin{equation}\label{SymDef}
 \left.
 Z {T}^{\dot{\alpha}}_{\alpha_1\dots\alpha_{2s-1}}\right|_{T_s=0}=0,
\end{equation}
where $T_s=0$ is a shorthand notation for (\ref{FEquations}). The
last relation means that the variation of the f\/ield equations
(\ref{FEquations}) along $Z$ is given by a linear combination of
these equations and their dif\/ferential consequences. In other words,
the local function of f\/ields $Z T_s$ vanishes on every solution to
the f\/ield equations. A symmetry is called \textit{trivial} if
\[
\left.Z_{\alpha_1\ldots\alpha_{2s}}\right|_{T_s=0}=0.
\]
All the symmetries below are considered modulo trivial ones.

A symmetry $Z$ is called {\textit{elementary}} if the functions
$Z_{\alpha_1\ldots\alpha_{2s}}$ are independent of f\/ields, i.e.,
they only depend on~$x$'s. Elementary symmetries correspond to
shifts of the f\/ield $\varphi$ by a particular solution of the
equations of motion. Clearly, such symmetries have a place in any
linear theory.

In \cite{AP2}, it was shown that all non-elementary symmetries of
equations (\ref{FEquations}) are generated by variational vector
f\/ields (\ref{RSymmetries}) whose coef\/f\/icients
$Z_{\alpha_1\ldots\alpha_{2s}}$ can be chosen to be linear in f\/ields
and their derivatives. For this reason we refer to these symmetries
as linear and denote the space of all linear symmetries by
$\mathrm{Sym}(T_s)$. There is an increasing f\/iltration of
$\mathrm{Sym}(T_s)$ by the subspaces of linear symmetries of order
$p$, namely,
\begin{equation}\label{SymFilt}
0=\mathrm{Sym}_{-1}(T_s) \subset \mathrm{Sym}_0(T_s)\subset
\mathrm{Sym}_1(T_s)\subset\cdots\subset
    \mathrm{Sym}_{\infty}(T_s)=\mathrm{Sym}(T_s) .
\end{equation}
It turns out that $\dim \mathrm{Sym}_p(T_s)<\infty$ for all
$p=0,1,\ldots$. The generators of symmetry transformations $Z$
def\/ine (and are def\/ined by) some linear endomorphisms $\hat{Z}$ of
the space of f\/ields. Linearity of the f\/ield equations
(\ref{FEquations}) and the condition~(\ref{SymDef}) suggest that
$\hat{Z}$ are precisely those endomorphisms that are interchangeable
with the wave operator $\hat{T}_s$ of equations~(\ref{FEquations})
in the following sense:
\[
\hat{T}_s  \hat{Z} = \hat{A}  \hat{T}_s
\]
for some matrix dif\/ferential operator $\hat{A}$. As a result, the
space $\mathrm{Sym}(T_s)$ carries the structure of an associative
f\/iltered algebra\footnote{For a general theory of f\/iltered and
graded algebras we refer the reader to~\cite{NVO}.} with respect to
the composition of endomorphisms underlying the symmetries. The
corresponding product
\[
\ast : \ \mathrm{Sym}_p(T_s)\times \mathrm{Sym}_q(T_s)\rightarrow
\mathrm{Sym}_{p+q}(T_s)
\]
can be written as follows: If $Z'$ and $Z''$ are two linear
symmetries, then $Z'\ast Z''$ is a linear symmetry generated by the
variational vector f\/ield~(\ref{RSymmetries}) with components
\[
Z_{\alpha_1\dots\alpha_{2s}}= Z'
(Z''_{\alpha_1\dots\alpha_{2s}}).
\]
Associated to the f\/iltered algebra (\ref{SymFilt}) is the graded
algebra
\[
    \mathrm{GrSym}(T_s)=\bigoplus_{p=0}^\infty
\mathrm{Sym}^p(T_s) ,\qquad
\mathrm{Sym}^{p}(T_s)=\mathrm{Sym}_{p}(T_s)/\mathrm{Sym}_{p-1}(T_s) .
\]
Although the algebras $\mathrm{Sym}(T_s)$ and $\mathrm{GrSym}(T_s)$
are not isomorphic to each other, there is an isomorphism of
f\/iltered vector spaces
\begin{equation}\label{iso}
\mathrm{GrSym}(T_s)\simeq\mathrm{Sym}(T_s) ,
\end{equation}
where the f\/iltration in $\mathrm{GrSym}(T_s)$ is given by the
standard f\/iltration of a graded vector space:
\[
\mathrm{GrSym}_p(T_s) \subset \mathrm{GrSym}_{p+1}(T_s) ,\qquad
\mathrm{GrSym}_p(T_s)=\bigoplus_{k=0}^p \mathrm{Sym}^k(T_s) ,\qquad
p=0,1,\ldots .
\]
The isomorphism (\ref{iso}) is far from being canonical, but it
implies the equality
\[
\dim \mathrm{GrSym}_p(T_s)=\dim \mathrm{Sym}_p(T_s) .
\]

As for any associative algebra, we can turn $\mathrm{Sym}(T_s)$ into
a f\/iltered Lie algebra with respect to the $\ast$-commutator
\[
[Z_1,Z_2]= Z_1\ast Z_2-Z_2\ast Z_1 ,
\]
which is actually given by the commutator of variational vector
f\/ields. Whereas the closedness with respect to commutation is an
inherent property of the inf\/initesimal symmetry transformations, the
existence of an associative $\ast$-product on symmetries is a
special property of linear f\/ield equations.

\subsection{Conserved currents and characteristics}

A \textit{conserved current} of order $p$ is a real vector-function
$ \Phi^{\alpha\dot\alpha} $ that depends on the $p$th jet of the
f\/ield $\varphi$ and satisf\/ies the condition
\[
    \left.
    \partial_{\alpha\dot\alpha}\Phi^{\alpha\dot\alpha}\right|_{T_s=0}=0 .
\]
A conserved current $\Phi$ is called \textit{trivial} if, being
evaluated on solutions to the f\/ield equations, it is given by the
divergence of some bivector f\/ield, that is,
\begin{equation}\label{TrCur}
    \Phi^{\alpha\dot\alpha}=
    \partial_{\beta}^{\dot\alpha}{\Theta}^{\alpha\beta}+ \mbox{c.c.} \qquad
    (\mathrm{mod} \  T_s)
\end{equation}
for some local functions of f\/ields
${\Theta}^{\alpha\beta}=\Theta^{\beta\alpha}$. The trivial currents,
having zero charge, are of no physical importance. Therefore, we
consider the conserved currents modulo trivial ones. It is easy to
see that in each equivalence class of a conserved current there is a
representative obeying the relation
\begin{equation}\label{QF}
    \partial_{\alpha\dot{\alpha}}\Phi^{\alpha\dot{\alpha}}=
    T^{\dot
    \alpha}_{\alpha_1\dots\alpha_{2s-1}}{Q}_{\dot{\alpha}}^{\alpha_1\dots\alpha_{2s-1}}+
    \mbox{c.c.},
\end{equation}
where ${Q}_{\dot{\alpha}}^{\alpha_1\dots\alpha_{2s-1}}$ is a local
function of f\/ields. The spinor-valued function ${Q}$ is called the
\textit{charac\-teristic} of the conserved current~$\Phi$. We say that
a characteristic is of order $p$ if it depends on the $p$th jet of
the f\/ield $\varphi$.

The existence of the Noether identities (\ref{NIdentities}) gives
rise to a great number of characteristics of the form
\[
    {Q}_{\dot{\alpha}}^{\alpha_1\dots\alpha_{2s-1}}=
    \partial^{(\alpha_1}_{\dot\alpha}{\chi}^{\alpha_2\dots\alpha_{2s-1})} ,
\]
where $\chi$ is an arbitrary symmetric spin-tensor f\/ield. All these
characteristics correspond to trivial conserved currents~(\ref{TrCur}) and can be ignored. Another source of triviality is
characteristics that are proportional to the (dif\/ferential
consequence of) f\/ield equations. Such characteristics also result in
trivial conserved currents and should be regarded as trivial. Taking
the quotient of the space of all characteristics by the subspace of
trivial ones, we get the space of nontrivial characteristics.

Actually, relation (\ref{QF}) gives rise to a one-to-one
correspondence between the spaces of nontrivial conserved currents
and characteristics, when either is properly def\/ined. For the
non-degenerate systems of PDEs such an isomorphism has been known
for a long time, see e.g.~\cite{Olver}. Its extension to the
Lagrangian  gauge systems was given in~\cite{BBH}, using some
previous results on the local BRST cohomology~\cite{DVHTV}. A
further generalization to the case of arbitrary (i.e., not
ne\-ces\-sarily Lagrangian and/or non-degenerate) system of PDEs subject
to the standard regularity conditions \cite{BBH} was formulated in
our recent paper~\cite{KLS3}. Thus, for regular PDEs the study of
conservation laws amounts to the study of characteristics and vice
versa. For the Bargmann--Wigner equations (\ref{FEquations}) this
general isomorphism was traced explicitly in \cite{AP1}.

In \cite{AP2}, it was shown that all the nontrivial characteristics
for equations (\ref{FEquations}) can be chosen to be at most linear
in f\/ields. Correspondingly, all the nontrivial conserved currents
can be chosen to be at most quadratic in f\/ields.  Below we are
restricted to the characteristics with linear dependence of f\/ields.
These form the linear space $\mathrm{Char}(T_s)$ f\/iltered by the
increasing sequence of f\/inite dimensional subspaces
$\mathrm{Char}_p(T_s)$ constituted by the characteristics of order~$p$,
\[
0=\mathrm{Char}_{-1}(T_s)\subset \mathrm{Char}_{0}(T_s)\subset
\mathrm{Char}_1(T_s)\subset\cdots\subset\mathrm{Char}_{\infty}(T_s)=\mathrm{Char}(T_s).
\]
Taking the successive quotients
\[
    \mathrm{Char}^p(T_s)=\mathrm{Char}_p(T_s)/\mathrm{Char}_{p-1}(T_s),
\]
we def\/ine  the associated graded space of linear characteristics
\[
\mathrm{GrChar}(T_s)=
\bigoplus_{p=0}^{\infty}\mathrm{Char}^{p}(T_s).
\]
As with the symmetries, there is an isomorphism of f\/iltered vector
spaces
\[
\mathrm{GrChar}(T_s)\simeq \mathrm{Char}(T_s)
\]
with respect to the canonical f\/iltration in $\mathrm{GrChar}(T_s)$,
\[
\mathrm{GrChar}_p(T_s) \subset \mathrm{GrChar}_{p+1}(T_s),\qquad
\mathrm{GrChar}_p(T_s)=\bigoplus_{k=0}^p\mathrm{Char}^p(T_s).
\]
In particular,
\[
\dim \mathrm{GrChar}_p(T_s)=\dim \mathrm{Char}_p(T_s).
\]

Characteristics form a f\/iltered module over the f\/iltered Lie algebra
$\mathrm{Sym}(T_s)$. The action of symmetries on characteristics
\[\odot
: \ \mathrm{Sym}_p(T_s)\times\mathrm{Char}_q(T_s)\rightarrow\mathrm{Char}_{p+q}(T_s)\]
can be def\/ined as follows. Let $Z$ be a symmetry and let $\Phi$ be
the conserved current associated to a characteristic $Q$. Then
$\Phi'=Z \Phi$ is a new conserved current and we def\/ine $Q'=Z\odot
Q$ to be the characteristic corresponding to $\Phi'$. Due to the
one-to-one correspondence between the spaces of nontrivial currents
and nontrivial characteristics the above def\/inition is consistent.
It should be noted that the space $\mathrm{Char}(T_s)$, being a
module over the Lie algebra $\mathrm{Sym}(T_s)$, is by no means  a
module over the associative algebra $(\mathrm{Sym}(T_s),\ast)$.

\section{Classif\/ication results}

As was mentioned in the Introduction a complete classif\/ication of
the rigid symmetries and conservation laws for purely bosonic f\/ields
(\ref{FEquations}) of spin $s\geq 1/2$ was given by Anco and
Pohjanpelto in~\cite{AP1,AP2} (see also~\cite{GSV}). Below we give a
brief summary of these results and comment on a~dif\/ference between
the bosonic and fermionic cases. An important part of the
classif\/ication is played by the algebra of Killing spinors on the
four-dimensional Minkowski space. Therefore we start with recalling
some basic facts concerning this algebra.

\subsection{The Killing spinors}

By def\/inition, a \textit{Killing spinor} is a spin-tensor f\/ield
$\xi^{\alpha_1\ldots\alpha_k}_{\dot\alpha_1\ldots\dot\alpha_l}(x)$
of type $(k,l)$ which is symmetric in dotted and undotted indices
and obeys the equation
\[
    \partial_{(\dot\alpha}^{(\alpha}\xi^{\alpha_1\ldots\alpha_k)}_{\dot\alpha_{1}\ldots\dot\alpha_l)}=0 .
\]
As above, the round brackets mean symmetrization of the enclosed
indices.

Let $\mathrm{Kil}(k,l)$ denote the space of all Killing spinors of
type $(k,l)$. The Killing spinors form a bigraded, associative,
commutative algebra $\mathrm{Kil}=\bigoplus_{k,l\in \mathbb{N}}
\mathrm{Kil}(k,l)$ with respect to the symmetrized tensor product
\[
\bullet: \ \mathrm{Kil}(k,l)\times \mathrm{Kil}(k',l')\rightarrow
\mathrm{Kil}(k+k',l+l') .
\]
If $\xi\in\mathrm{Kil}(k,l)$ and $\xi'\in\mathrm{Kil}(k',l')$ then
$\xi\circ \xi'$ is the Killing spinor of type $(k+k',l+l')$ with
components
\begin{equation}\label{stp}
    (\xi\bullet \xi')^{\alpha_1\ldots\alpha_{k+k'}}_{\dot{\alpha}_1\ldots\dot{\alpha}_{l+l'}}
    =\xi^{(\alpha_1\ldots\alpha_{k}}_{(\dot{\alpha}_1\ldots\dot{\alpha}_{l}}
    \xi'^{\alpha_{k+1}\ldots\alpha_{k+k'})}_{\dot{\alpha}_{l+1}\ldots\dot{\alpha}_{l+l'})} .
\end{equation}
The complex conjugation of spin-tensor f\/ields, $\xi\mapsto \bar\xi$,
def\/ines an involution in the algebra $\mathrm{Kil}$ such that
$\overline{\mathrm{Kil}(k,l)}\simeq \mathrm{Kil}(l,k)$. The Killing
spinor $\xi$ of type $(s,s)$ is called \textit{real} if $\bar \xi
=\xi$. The space of all real spinors of type $(s,s)$ will be denoted
by $\mathrm{Kil}_{\mathbb{R}}(s,s)$.

In what follows we will be mostly interested in Killing spinors
belonging to the following sequence of vector spaces
\[
    \mathrm{Kil}_{s}=\bigoplus_{p=0}^{\infty}\mathrm{Kil}^p_s,\qquad\mathrm{Kil}^p_s=\mathrm{Kil}(p,p)\oplus
    \mathrm{Kil}(p-2s,p+2s),\qquad
    s=\frac{1}{2},1,\frac{3}{2},2, \ldots .
\]
Here it is implied that $\mathrm{Kil}(p-2s,p+2s)=0$ whenever $2s>p$.
It is well known that each homogeneous subspace of $\mathrm{Kil}_s$
is of f\/inite dimension, namely,
\[
\dim_\mathbb{C} \mathrm{Kil}(p-2s,p+2s)=
(p-2s+1)(p-2s+2)(p+2s+1)(p+2s+2)(2p+3)/12 .
\]

Although the  space $\mathrm{Kil}_s$ is not a subalgebra with
respect to the $\bullet$-product (\ref{stp}),  it can be endowed
with the structure of an associative \textit{superalgebra} with
respect to a new product
\[
\circ: \ \mathrm{Kil}^p_s\times \mathrm{Kil}^q_s\rightarrow
\mathrm{Kil}^{p+q}_s .
\]
The $\mathbb{Z}_2$-grading on $\mathrm{Kil}_s$ is def\/ined by the
direct sum decomposition
\[
\mathrm{Kil}_s=\mathrm{Kil}^+_s\oplus \mathrm{Kil}^-_s ,
\]
 where
\[
\mathrm{Kil}_s^+=\bigoplus_{p=0}^\infty \mathrm{Kil}(p,p) ,\qquad
\mathrm{Kil}_s^-=\bigoplus_{p=0}^\infty \mathrm{Kil}(p,p+4s) .
\]
Then for all $\xi,\xi'\in \mathrm{Kil}^+_s$ and $\eta,\eta'\in
\mathrm{Kil}^-_s$ we set
\[
\xi\circ \xi'=\xi\bullet \xi' ,\qquad
\xi\circ\eta=\bar\xi\bullet\eta ,\qquad
\eta\circ\xi=\eta\bullet\xi ,\qquad
\eta\circ\eta'=\bar\eta\bullet\eta' .
\]
One can easily verify that, being def\/ined in such a way, the
$\circ$-product is associative, though not commutative. At the same
time, its restriction to the even subalgebra $\mathrm{Kil}_s^+$
coincides with the $\bullet$-product and is thus commutative.
Involving the complex conjugation, the $\circ$-product is only
$\mathbb{R}$-linear, not $\mathbb{C}$-linear.

As an associative algebra the subalgebra $\mathrm{Kil}^+_s$ is
generated by the Killing spinors of type~$(1,1)$, so that each even
element can be represented as a sum of elements of the form
\begin{equation}\label{b1}
\xi_1\circ \xi_2\circ\cdots\circ\xi_p ,\qquad \xi_i\in
\mathrm{Kil}(1,1) .
\end{equation}
Similarly, each element from the odd subspace $\mathrm{Kil}^-_s$ is
given by a linear combination of the Killing spinors
\begin{equation}\label{b2}
\xi_1\circ\xi_2\circ\cdots\circ\xi_p\circ\Upsilon ,\qquad
\Upsilon\in \mathrm{Kil}(0,4s) .
\end{equation}
Having in mind the standard relation between tensor and spin-tensor
f\/ields  in four-dimensional Minkowski space,  we refer to type
$(p,p)$ Killing spinors as \textit{Killing tensors} of rank $p$ and
to type $(0,4s)$ Killing spinor as  self-dual Killing--Yano tensors
of rank~$2s$. The Killing tensors of rank $1$ are just the Killing
vectors.

\subsection{Classif\/ication}

A general
correspondence between the space of Killing spinors, nontrivial
symmetries and cha\-rac\-teristics can be summarized by the diagram of
maps
\[
\xymatrix{&\mathrm{Kil}_s\ar[dl]_{\pi_s}\ar[dr]^{\sigma_{s}}&\\
\mathrm{GrChar}(T_s)& & \mathrm{GrSym}(T_s)}
\]
with the following properties:
\begin{enumerate}\itemsep=0pt
    \item The homomorphisms $\pi_s$ and
$\sigma_s$ are homogeneous of degrees $1-2s$ and $0$, respectively.
Hence,
\[
\pi_s: \ \mathrm{Kil}_s^p\rightarrow \mathrm{Char}^{p-2s+1}
(T_s) ,\qquad \sigma_s: \ \mathrm{Kil}_s^p\rightarrow
\mathrm{Sym}^p(T_s)  .
\]
    \item $\sigma_s$ is an isomorphism of graded associative algebras.

    \item $\pi_s$ is a surjection.
\end{enumerate}

Let us f\/irst describe the homomorphism  $\sigma_s$. By def\/inition,
it takes the generators
\begin{equation}\label{mgen}
1\in \mathrm{Kil}(0,0) ,\qquad \xi\in \mathrm{Kil}(1,1),\qquad
\Upsilon\in \mathrm{Kil}(0,4s)
\end{equation}
of the algebra $\mathrm{Kil}_s$ to the variational vector f\/ield
(\ref{RSymmetries}) with components
\begin{gather*} Z[1]_{\alpha_1\dots \alpha_{2s}}=\varphi_{\alpha_1\dots
\alpha_{2s}},\nonumber\\
    Z[\xi]_{\alpha_1\ldots\alpha_{2s}}=\xi^{\beta\dot{\beta}}\partial_{\beta\dot\beta}\varphi_{\alpha_1\ldots\alpha_{2s}}+
    s\partial_{(\alpha_1\dot{\beta}}\xi^{\beta\dot{\beta}}\varphi_{\beta\alpha_2\ldots\alpha_{2s})}+
    \frac{1-s}{4}\left(\partial_{\beta\dot{\beta}}\xi^{\beta\dot{\beta}}\right)\varphi_{\alpha_1\ldots\alpha_{2s}}, \\
    Z[\Upsilon]_{\alpha_1\ldots\alpha_{2s}}=\sum_{p=0}^{2s}\frac{2s-p+1}{4s+1} \binom{2s}{p}\left(\partial_{(\alpha_{1}}^{\dot\beta_1}
    \cdots\partial_{\alpha_{p}}^{\dot\beta_p}\Upsilon_{\dot\beta_1\dots\dot\beta_{4s}}\right)
    \left(\partial^{\dot\beta_{p+1}}_{\alpha_{p+1}}\cdots\partial^{\dot\beta_{2s}}_{\alpha_{2s})}
    \bar\varphi^{\dot\beta_{2s+1}\dots \dot\beta_{4s}}\right).
\end{gather*}

Clearly, $Z[1]$ generates the dilatations of the f\/ield $\varphi$ and
$Z[\xi]$ corresponds to the ``conformally weighted'' Lie derivative
of $\varphi$ along the Killing vector $\xi$. By property~(2) the
homomorphism $\sigma_s$ is uniquely extended from the multiplicative
(\ref{mgen}) to linear generators~(\ref{b1}), (\ref{b2}) of the
algebra $\mathrm{Kil}_s$:
\begin{gather}
Z[\xi_1\circ\cdots\circ\xi_p]=Z[\xi_1]\ast\cdots\ast
    Z[\xi_p] ,\nonumber\\
    Z[\zeta_1\circ\cdots\circ\zeta_{p-2s}\circ\Upsilon]=Z[\zeta_1]\ast\cdots\ast
    Z[\zeta_{p-2s}]\ast Z[\Upsilon] .\label{KiltoSym}
\end{gather}
Hereafter we use the Greek letter $\zeta$ to denote the real Killing
vectors, i.e., elements of $\mathrm{Kil}_{\mathbb{R}}(1,1)$, while~$\xi$ is used for complex Killing vectors. Notice that the
symmetries in the right hand sides of~(\ref{KiltoSym}) represent
elements of the graded vector space $\mathrm{GrSym}(T_s)$ and, as
elements of $\mathrm{GrSym}(T_s)$, are completely symmetric in
permutations of~$\xi$'s and~$\zeta$'s. The last fact easily follows
from the commutation relations
\[
[Z[\xi_1],Z[\xi_2]]=Z[[\xi_1,\xi_2]],\qquad
    [Z[\Upsilon],Z[\zeta]]=Z[\mathcal{L}_\zeta\Upsilon] .
\]

The symmetries (\ref{KiltoSym}) exhaust all the non-elementary
symmetries of the relativistic wave equation  (\ref{FEquations}).
The space of symmetries inherits the $\mathbb{Z}_2$-grading from the
space of Killing tensors, namely, the even symmetries take the f\/ield
to f\/ield, while the odd symmetries take the  f\/ield to its complex
conjugate.

Let us now turn to the homomorphism~$\pi_s$. Again, the
$\mathbb{Z}_2$-grading on the space $\mathrm{Kil}_s$ induces the
same grading on the space of characteristics.  As a
$\mathbb{Z}_2$-graded module over the Lie algebra of symmetries the
space~$\mathrm{GrChar}(T_s)$ is generated by the even
characteristics of order zero
\begin{equation}\label{CharT0}\displaystyle
    Q[i^{2s}\zeta_{1}\circ\cdots\circ\zeta_{2s-1}]_{\alpha}^{\dot{\alpha}_1\dots\dot{\alpha}_{2s-1}}=
    (\zeta_{1}\circ\cdots\circ\zeta_{2s-1})^{\alpha_1\dots\alpha_{2s-1}\dot{\alpha}_1\dots\dot{\alpha}_{2s-1}}
    \varphi_{\alpha\alpha_1\dots\alpha_{2s-1}}
\end{equation}
and  some  odd characteristics of order one or two depending on $s$.
For half-integer spins the latter characteristics are given by
\begin{gather}
    Q[i^{2s}\Upsilon]_{\alpha}^{\dot{\alpha}_1\dots\dot{\alpha}_{2s-1}}=
    \Upsilon^{\dot{\alpha}\dot{\alpha}_1\dots\dot{\alpha}_{2s-1}\dot{\beta_1}\dots\dot{\beta}_{2s}}\partial_{\alpha\dot{\alpha}}
    \bar{\varphi}_{\dot{\beta}_1\dots\dot{\beta}_{2s}}\nonumber\\
\phantom{Q[i^{2s}\Upsilon]_{\alpha}^{\dot{\alpha}_1\dots\dot{\alpha}_{2s-1}}=}{}
+\frac{2s+1}{4s+1}\partial_{\alpha\dot{\alpha}}
    \Upsilon^{\dot{\alpha}\dot{\alpha}_1\dots\dot{\alpha}_{2s-1}\dot{\beta_1}\dots\dot{\beta}_{2s}}
    \bar{\varphi}_{\dot{\beta}_1\dots\dot{\beta}_{2s}}\label{QY}
\end{gather} and for integer spins they can be written as
\begin{equation}\label{V1}
    Q[\zeta\circ\Upsilon]=Z[\zeta]Q[\Upsilon]+\frac{1}{2}Q[\mathcal{L}_\zeta\Upsilon].
\end{equation}
Now the space of characteristics for real-valued conserved currents
is spanned by the even charac\-teristics
\begin{equation}\label{CharT}
    Q[i^{q+2s}\zeta_{1}\circ\cdots\circ\zeta_{q+2s-1}]=Z[i^q\zeta_{1}\circ\cdots\circ\zeta_{q}]\odot
    Q[i^{2s}\zeta_{q+1}\circ\cdots\circ\zeta_{q+2s-1}]
\end{equation}
and odd characteristics
\begin{alignat}{3}
& Q[\zeta_{1}\circ\cdots\circ\zeta_{2p}\circ \Upsilon]=
    Z[\zeta_{1}\circ\cdots\circ\zeta_{2p}]\odot
    Q[\Upsilon], \qquad  && s\in \mathbb{N}-\frac12 ,& \nonumber\\
&    Q[\zeta_{1}\circ\cdots\circ\zeta_{2p+1}\circ\Upsilon]=
    Z[\zeta_{1}\circ\cdots\circ\zeta_{2p}]\odot
    Q[\zeta_{2p+1}\circ\Upsilon] , \qquad && s\in \mathbb{N}  . & \label{CharV}
\end{alignat}
The characteristics (\ref{CharT}), (\ref{CharV}) have the orders
$q$,  $2p+1$ and $2p+2$, respectively. Let us stress again that the
expressions in the right hand sides of (\ref{CharT}) and
(\ref{CharV}) are considered as representing elements of
$\mathrm{GrChar}(T_s)$, i.e., modulo characteristics of lower
orders. Upon this interpretation the right hand sides are totally
symmetric in $\zeta$'s.

Formulae (\ref{CharT0})--%, (\ref{QY}), (\ref{V1}), (\ref{CharT}),
(\ref{CharV}) provide a complete description of the homomorphism
$\pi_s$. In particular, they show that $\pi_s$ is surjective. The
kernel of $\pi_s$ is given by the space
\[
\mathrm{Ker}\,\pi_s=\mathrm{Ker}_+\pi_s\oplus \mathrm{Ker}_-\pi_s ,
\]
where
\[
\mathrm{Ker}_+\pi_s =\left(\bigoplus_{n=0}^{2s-2}\mathrm{Kil}(n,n)
\right)\oplus \left(\bigoplus_{m=2s-1}^\infty i^{m}
\mathrm{Kil}_\mathbb{R}(m,m) \right)
\]
and
\[
\mathrm{Ker}_{-}\pi_s=\left\{
\begin{array}{ll}
   \displaystyle \bigoplus_{p=0}^\infty \mathrm{Kil}(2p+1, 2p+4s+1) , & \hbox{for $s\in \mathbb{N}-\frac12$;} \\[5mm]
   \displaystyle \bigoplus_{p=0}^\infty \mathrm{Kil}(2p,2p+4s), & \hbox{for $s\in \mathbb{N}$.} \\
\end{array}
\right.
\]

Introduce the subspace ${\mathrm{ChKil}}_s\subset {\mathrm{Kil}}_s$
which is complementary to $\mathrm{Ker}\,\pi_s$:
\[
{\mathrm{ChKil}}_s={\mathrm{ChKil}}_s^+\oplus
{\mathrm{ChKil}}_s^- ,
\]
where
\[
{\mathrm{ChKil}}_s^+=\bigoplus_{m=2s-1}^\infty
i^{m+1}\mathrm{Kil}_\mathbb{R}(m,m)
\]
and
\[
{\mathrm{ChKil}}_s^- = \left\{
\begin{array}{ll}
   \displaystyle \bigoplus_{p=0}^\infty \mathrm{Kil}(2p, 2p+4s), & \hbox{for $s\in \mathbb{N}-\frac12$;} \\[5mm]
   \displaystyle \bigoplus_{p=0}^\infty \mathrm{Kil}(2p+1,2p+4s+1), & \hbox{for $s\in \mathbb{N}$.} \\
\end{array}
\right.
\]
With the def\/initions above we have the direct sum decomposition
\[
\mathrm{Kil}_s=\mathrm{Ker}\,\pi_s\oplus{\mathrm{ChKil}}_s
\]
and $\pi_s$ becomes an isomorphism when restricted to
${\mathrm{ChKil}}_s$. So, we can draw the following diagram:
\begin{equation}\label{XYpp}
\xymatrix{&\mathrm{ChKil}_s\ar[dl]_{{\pi}_s}\ar[dr]^{{\sigma}{}_{s}}&\\
\mathrm{GrChar}(T_s)& & \mathrm{GrSym}(T_s)}
\end{equation}
where $\sigma_s$ is sill an injection.
  For an obvious reason we
refer to $\mathrm{ChKil}_s$ as the space of \textit{characteristic
Killing tensors}. As is seen, no characteristics are assigned to the
Killing tensors of rank $<2s-1$ and only the odd (even) rank tensors
of Killing and Yano correspond to characteristics for integer
(half-integer) spins.

\begin{remark} The original
classif\/ication \cite{AP1}  of the quadratic conserved currents was
formulated for purely bosonic f\/ields. It essentially relied on the
notion of an \textit{adjoint symmetry}, see equation~(\ref{AP}) of
Appendix~\ref{appendixA}. Each Killing spinor $\xi\in\mathrm{Kil}_s^p$, $p\geq
2s-1$, was shown to give rise to an adjoint symmetry and a conserved
current. By construction, such currents exhaust all the quadratic
conserved currents, but dependencies are allowed (i.e., some of
nontrivial linear combinations of the conserved currents may give a
trivial current). Actually, only a ``half'' of the Killing spinors
above was shown to generate a basis of the nontrivial conserved
currents \cite[Corollary~4.4]{AP1}. This classif\/ication method is
directly applicable to the fermionic f\/ields as well, but the
resulting conserved currents are dif\/ferent. For the fermionic f\/ields
of half-integer spin, exactly the complementary ``half'' of the
Killing spinors spans the space of nontrivial quadratic currents.
Expressions~(\ref{CharT}),~(\ref{CharV}) are given for the standard
spin-statistics correspondence. Another approach to the construction
of conserved currents for massless f\/ields has been proposed in~\cite{GSV}. In that paper, the conserved currents have been def\/ined
as contractions of \textit{conserved tensors} with appropriate
Killing spinors, so that the map $\pi_s$ has appeared in quite a
natural way. However, the dependencies have been unnoticed between
the currents, and the kernel of $\pi_s$ was not studied. With a due
account of these dependencies and the statistics, the real-valued
conserved currents found in~\cite{GSV} coincide with those
constructed and classif\/ied in~\cite{AP1}.
\end{remark}

\section{Lagrange anchor and characteristic symmetries}\label{CS}

The Lagrange anchor was f\/irst introduced in \cite{KazLS} as a tool
for the path-integral quantization of non-Lagrangian theories.
Later, it has been realized \cite{KLS2} that the concept of Lagrange
anchor can also serve in classical theory for establishing a
systematic correspondence between conservation laws and symmetries,
providing a generalization of the f\/irst Noether's theorem to
non-Lagrangian dynamics. The latter result admits also a natural
cohomological interpretation within the BRST formalism \cite{KLS3}.
In the body of the paper, we do not elaborate on the def\/inition and
general properties of the Lagrange anchor, which are explained from
various viewpoints in the cited works. A brief account of the
general notions concerning the Lagrange anchor can be found in
Appendix~\ref{appendixA}.

In this particular model, one can arrive at the Lagrange anchor
proceeding from the following simple observation. Since the map
${\pi}_s: \mathrm{ChKil}_s\rightarrow \mathrm{GrChar}(T_s)$ is an
isomorphism, the diagram~(\ref{XYpp}) can be completed uniquely to
the commutative triangle diagram
\begin{equation}\label{XYppp}
\xymatrix{&\mathrm{ChKil}_s\ar[dl]_{{\pi}_s}\ar[dr]^{{\sigma}{}_{s}}&\\
\mathrm{GrChar}(T_s)\ar[rr]^{V_s={\sigma}_s {\pi}_{s}^{-1}}& &
\mathrm{GrSym}(T_s)}
\end{equation}
Now, one can see that the bottom map is given by a universal linear
dif\/ferential operator acting from the space of characteristics to
the space of symmetries. It is the operator that can be taken as a
Lagrange anchor for the f\/ield equations~(\ref{FEquations}).
Explicitly, if $Q=(Q^{\alpha_1\dots\alpha_{2s-1}}_{\dot \alpha})$
is a~characteristic, then $V_s$ takes it to the symmetry~(\ref{RSymmetries}) with
\begin{equation}\label{V}
Z_{\alpha_1\dots\alpha_{2s}}=
V_s(Q)_{\alpha_1\dots\alpha_{2s}}=i^{2s}\partial_{(\alpha_2\dot\alpha_2}\dots\partial_{\alpha_{2s-1}\dot\alpha_{2s-1}}
    \bar Q_{\alpha_1)}^{\dot\alpha_2\dots\dot\alpha_{2s}} .
\end{equation}
Verif\/ication of the def\/ining property of a Lagrange anchor
(\ref{AVT}) is straightforward, see Appendix~\ref{appendixB}. Actually, formula
(\ref{V}), considered for particular representatives of
characteristics and symmetries, def\/ines the map
\begin{equation}\label{Vhom}
    V_s: \ \mathrm{Char}_{p}(T_s)\rightarrow
    \mathrm{Sym}_{p+2s-1}(T_s) ,\qquad p=0,1,2, \ldots,
\end{equation}
of the underlying f\/iltered spaces, from which the induced map
(\ref{XYppp}) of the associated graded spaces follows. (By abuse of
notation, we denote both the maps by~$V_s$.) The map~(\ref{Vhom}),
being def\/ined through the composition of monomorphisms~$\sigma_s$
and $\pi^{-1}_s$, is obviously injective and we denote its image by
$\mathrm{ChSym}_p(T_s)$. The space
$\mathrm{ChSym}(T_s)=\mathrm{ChSym}_{\infty}(T_s)$ is referred to as
the space of \textit{characteristic symmetries}.

Being independent of f\/ields, the Lagrange anchor $V_s$ is
automatically \textit{strongly integrable}, see relations~(\ref{AAn}) of
Appendix~\ref{appendixA}. The last fact implies that the characteristic symmetries
form a~subalgebra in the Lie algebra~$\mathrm{Sym}(T_s)$. We have
\begin{equation}\label{Hom}
[V_s(Q_1),V_s(Q_2)] = V_s([Q_1,Q_2]_{V_s}) , \qquad \forall\,
Q_1,Q_2\in \mathrm{Char}(T_s) ,
\end{equation}
where
\[
[Q_1,Q_2]_{V_s}:=V_s(Q_1)\odot Q_2=-V_s(Q_2)\odot Q_1 .
\]
The skew-symmetric bracket operation
\[
[\;\cdot\;,\;\cdot\;]_{V_s}: \ \mathrm{Char}(T_s)\times\mathrm{Char}(T_s)\rightarrow\mathrm{Char}(T_s)
\]
enjoys the Jacobi identity making $\mathrm{Charq}(T_s)$ into a Lie
algebra. This is a particular case of the Lie bracket on
characteristics introduced in \cite{KLS2}\footnote{For Lagrangian
theories endowed with the \textit{canonical} Lagrange anchor
\cite{KazLS} this bracket reproduces the well-known Dickey's bracket
on conserved currents~\cite{D} (see also~\cite{BH}).}. Equation~(\ref{Hom}) says the map (\ref{Vhom}) def\/ines a homomorphism from
the Lie algebra of characteristics to the Lie algebra of
characteristic symmetries. For low spins ($s=1/2, 1$) the Lie
algebra $\mathrm{Char}(T_s)$ contains a f\/inite dimensional subalgebra
which is isomorphic to the Lie algebra of conformal group. The
elements of this subalgebra correspond to conserved currents that
are expressible in terms of the energy-momentum tensor.

\begin{remark} The map (\ref{Vhom})
extends naturally from the space of characteristics to the space of
all adjoint symmetries, keeping the same value area. Actually, this
is a general property of any system of PDEs, and not just a special
feature of the Bargmann--Wigner equations (see relation~(\ref{AV}) in
Appendix~\ref{appendixA}). If one includes the adjoint symmetries of~(\ref{FEquations}) to the range of def\/inition, then the image of
$V_s$ will cover all the symmetries of the system, except for a
f\/inite number of symmetries of order $\leq 2s-1$. Although~(\ref{AV}) has the form of a linear mapping, it should be better
thought of as a universal bilinear map assigning a symmetry~$Z$ to
any pair $(V,P)$ constituted by a Lagrange anchor~$V$ and an adjoint
symmetry $P$. The map is universal as it does not depend on a
particular structure of the f\/ield equations~(\ref{AT}). In this
context it is pertinent to mention another universal bilinear map
connecting symmetries, adjoint symmetries, and characteristics
\cite{AB1,BCA}. This takes a symmetry~$Z$ and an adjoint symmetry~$P$ to a characteristic~$Q$. For each specif\/ic choice of~$P$ the
assignment $(Z,P)\mapsto Q$ def\/ines a linear mapping from the space
of symmetries to the space of characteristics. This map, however,
acts in the opposite direction with respect to the Lagrange anchor
map~(\ref{AV}). Either of the universal mappings can be regarded as
an extension of the classical Noether's theorem to non-variational
PDEs, though the constructions are not entirely peer entities, in
the following sense. For Lagrangian equations, there is a preferable
choice for the Lagrange anchor that immediately reproduces the
standard Noether's correspondence between symmetries and
conservation laws. It is unlikely that such a canonical choice can
exist for the adjoint symmetry of Lagrangian equations.
\end{remark}

\section[Probability amplitude for massless fields of spin $s\geq 1/2$]{Probability amplitude for massless f\/ields of spin $\boldsymbol{s\geq 1/2}$}

In this section, we brief\/ly comment on how the Lagrange anchor above
can be used to perform a consistent path-integral quantization of
the massless f\/ields subject to the Bargmann--Wigner equations.

Recall that in the covariant formulation of quantum f\/ield theory one
usually deals with the path integrals of the form
\begin{equation}\label{O}
    \langle \mathcal{O}\rangle_{\Psi}=\int
    [D\phi]\mathcal{O}[\phi]\Psi[\phi] ,
\end{equation}
where $\mathcal{O}$ is a physical observable and $\Psi$ is a
probability amplitude on the conf\/iguration space of f\/ields $\phi^i$.
For a Lagrangian theory with action $S[\phi]$ the latter is given by
the Feynman amplitude $\Psi=e^{\frac i{\hbar}S}$, which can also be
def\/ined as a unique (up to a normalization factor) solution to the
Schwinger--Dyson (SD) equation
\begin{equation}\label{S-D}
    \left(\frac{\delta S}{\delta \phi^i}+i\hbar
    \frac{\delta}{\delta\phi^i}\right)\Psi[\phi]=0 .
\end{equation}
After normalization, the integral (\ref{O}) def\/ines the quantum
average of a physical observable $\mathcal{O}$ in the theory with
probability amplitude $\Psi$. It is believed that evaluating such
integrals for various observables~$\mathcal{O}$ one can extract all
physically relevant information about the quantum dynamics of f\/ields
(e.g.\ the scattering matrix).

In \cite{LS1},  the SD equation (\ref{S-D}) was shown to admit quite
a natural generalization to non-Lagrangian theories endowed with
compatible Lagrange anchors. Leaving aside general de\-f\/i\-ni\-tions, we
just present the generalized SD equations for the f\/ield equations~(\ref{FEquations}) and  the Lagrange anchor~(\ref{V}). These read
\begin{gather}
\left({T}^{\dot{\alpha}}_{\alpha_1\dots\alpha_{2s-1}}-
i^{2s}\hbar\partial_{\alpha_1\dot\alpha_1}\cdots\partial_{\alpha_{2s-1}\dot\alpha_{2s-1}}\frac{\delta}{\delta
\bar\varphi_{\dot\alpha\dot\alpha_1\dots\dot\alpha_{2s-1}}}\right)
\Psi[\varphi]=0 ,\nonumber\\
 \left(\bar{T}^{{\alpha}}_{\dot\alpha_1\dots\dot\alpha_{2s-1}}-
i^{-2s}\hbar\partial_{\alpha_1\dot\alpha_1}\cdots\partial_{\alpha_{2s-1}\dot\alpha_{2s-1}}\frac{\delta}{\delta
\varphi_{\alpha\alpha_1\dots\alpha_{2s-1}}}\right)
\Psi[\varphi]=0 .\label{SD}
\end{gather}
As with the usual SD equation (\ref{S-D}), the f\/irst terms in
(\ref{SD}) are determined by the classical equations of motion~$T_s$
and the second ones, constructed by the Lagrange anchor~$V_s$,
involve the f\/irst-order variational derivatives multiplied by the
Plank constant. In the classical limit, $\hbar\rightarrow 0$, the
latter terms vanish and the resulting probability amplitude is given
by the Dirac distribution supported at the classical solutions to
the f\/ield equations. Formally, $\Psi|_{\hbar\rightarrow 0}\sim
\delta(T_s)$ and one can think of the last expression as a classical
probability amplitude. The generalized SD equations~(\ref{SD}) are
formally consistent since the linear operators determining the left
hand sides commute to each other\footnote{In fact, it is the
requirement of formal compatibility of the generalized SD equations
that determines all possible Lagrange anchors for given equations of
motion.}. The general solution to (\ref{SD}) admits a nice
path-integral representation in terms of the action functional for
the so-called \textit{augmented theory}. Within the augmentation
approach \cite{LS2} the original set of f\/ields
$\varphi_{\alpha_1\dots\alpha_{2s}}$ is extended by the new f\/ields
$Y^{\alpha_1\dots\alpha_{2s-1}}_{\dot\alpha}$ that take values in
the space dual to the space of equations of motion. The action of
the augmented theory is systematically constructed by the original
equations of motion and the corresponding Lagrange anchor. In the
case at hand we f\/ind
\begin{gather*}
 S_{\mathrm{aug}}[\varphi,Y]=\displaystyle\int d^4x \Big(
Y_{\dot\alpha}^{\alpha_1\dots\alpha_{2s-1}}{T}^{\dot\alpha}_{\alpha_1\dots\alpha_{2s-1}}
 \\
\hphantom{S_{\mathrm{aug}}[\varphi,Y]=\displaystyle\int d^4x \Big(}{}
+ {i^{2s}}{2}\partial_{\alpha_1(\dot\alpha_1}\cdots\partial_{\alpha_{2s-1}\dot\alpha_{2s-1}}
Y_{\dot\alpha_{2s})}^{\alpha_1\dots\alpha_{2s-1}}
\partial^{\alpha_{2s}\dot\alpha_1}{\bar{Y}}_{\alpha_{2s}}^{\dot\alpha_{2}\dots\dot\alpha_{2s}}+\mathrm{c.c.}\Big) .
\end{gather*}
The   least action principle results in the system of decoupled
equations
\[
\partial^{\alpha\dot{\alpha}}\varphi_{\alpha\alpha_1\dots\alpha_{2s-1}}=0 ,\qquad
\partial^{(\alpha_{1}\dot\alpha}Y_{\dot\alpha}^{\alpha_2\dots\alpha_{2s})}=0 .
\]
The dynamics of $\varphi$'s are seen to be governed by the original
equations of motion (\ref{FEquations}). The equations for the
augmentation f\/ields $Y$ are known as the {\textit{adjoint
equations}}. The latter are the starting point for the construction
of conserved currents by method~\cite{AP1}.

Integrating by parts under the path-integral sign, one can easily
verify that the functional
\begin{equation}\label{PA}
\Psi[\varphi]=N\int [DY]e^{\frac{i}{\hbar}S_{\mathrm{aug}}[\varphi,
Y]}
\end{equation}
does satisfy (\ref{SD}) for an arbitrary normalization constant $N$.
Doing the Gaussian integral (\ref{PA}), one can then f\/ind an
explicit expression for the functional $\Psi$. As an example,
consider the case $s=1/2$. The integral (\ref{PA}) takes the form
\[
\Psi[\varphi]=N\int [DY]\exp\left(\frac{i}{\hbar}\int d^4x
Y_{\dot\alpha}{T}^{\dot\alpha}-\bar{Y}_{\alpha}\bar{{T}}^\alpha+i{Y}_{\dot\alpha}
\partial^{\alpha\dot\alpha}\bar Y_{\alpha}\right).
\]
Making the substitution
$Y_{\dot\alpha}\mapsto{Y}_{\dot\alpha}-i\bar\varphi_{\dot\alpha}$
and integrating by $Y$'s,  we obtain
\[
\Psi[\varphi]= N'e^{\frac{i}{\hbar}S_{{1}/{2}}[\varphi]} ,
\]
where $N'$ is some constant and
\[
S_{{1}/{2}}[\varphi]=-i\int d^4x
\bar{\varphi}^{\dot\alpha}\partial_{\alpha\dot\alpha}\varphi^{\alpha}
\]
is the usual action for spin-$1/2$ massless f\/ield. Similar
computations\footnote{Notice that the Noether identities~(\ref{NIdentities}) give rise to certain gauge symmetries in the
augmented theory~\cite{LS2}, namely, $
{\delta_{\varepsilon}Y}_{\dot{\alpha}}^{\alpha_1\dots\alpha_{2s-1}}=
    \partial^{(\alpha_1}_{\dot\alpha}{\varepsilon}^{\alpha_2\dots\alpha_{2s-1})}$,  $\delta_{\varepsilon}\varphi=0$.  Therefore, to compute
the Gaussian integral (\ref{PA}) for $s>1/2$ one should f\/irst impose
an appropriate gauge f\/ixing condition on $Y$'s.} for $s>1/2$ result
in the probability amplitude $\Psi\sim e^{\frac i{\hbar}S}$, where
the exponent $S[\varphi]$ is no longer a local functional. An
explicit expression for the spin-$1$ non-local action $S_1[\varphi]$
can be found in \cite{LS2}.

Notice that the action of the augmented theory is manifestly
Poincar\'e invariant for any spin.  This suggests that the
relativistic symmetries presumably survive  quantization. It would
be interesting to establish which higher symmetries of the original
f\/ield equations can be ``lifted'' to the augmented theory and then
to quantum theory. Another interesting problem is to classify all
nontrivial Lagrange anchors for the Bargmann--Wigner equations
(\ref{FEquations}). This problem is in principle of the same kind as
the classif\/ication of symmetries or characteristics, though no
explicit  examples are known except for the Lagrange anchor
considered in the present paper. This seems to be a unique Lagrange
anchor enjoying the Poincar\'e invariance.

\section{Conclusion}

In this paper, we have found a Poincar\'e invariant Lagrange anchor
for the Bargmann--Wigner equations and applied this anchor for
deriving symmetries from conservation laws and for def\/ining the
quantum probability amplitude. Of course, at the level of free
f\/ields there is an equivalent Lagrangian formulation due to Fang and
Fronsdal~\cite{FF, F}. This formulation allows one to solve the above
problems by standard tools of Lagrangian f\/ield theory. The
Lagrangian formulation, however, does not admit consistent
interactions, while the non-Lagrangian equations for interacting
higher-spin f\/ields have been proposed by Vasiliev~\cite{V1,V2}. The
Lagrange anchor for Vasiliev's equations is still unknown. If the
anchor is found, it will allow one to develop a~consistent quantum
theory of higher-spin interactions. There are many other
non-Lagrangian models of physical interest for which the concept of
Lagrange anchor seems having no alternatives when it comes to
quantization or establishing a systematic connection between
symmetries and conservation laws. The procedure of f\/inding the
Lagrange anchor in the considered simple model is quite general and
it can be instructive in more complex theories.

\appendix

\section{General def\/inition and properties of a Lagrange anchor}\label{appendixA}

To make the paper self-contained, we explain here some basic notions
concerning the concept of Lagrange anchor. A more extended and
systematic exposition can be found in the original papers
\cite{KLS2,KazLS,LS2}. To emphasize the algebraic structure
underlying the concept, we use De Witt's condensed notation
\cite{DW}, which is much more handy and compact than the jet space
formalism, given the context. According to this notation the f\/ields
$\phi^i(x)$, leaving on a smooth manifold~$X$, are interpreted as
local coordinates on an inf\/inite dimensional supermanifold $M$. The
local coordinates $x$'s on $X$ are treated as continuous indices
labeling the f\/ields and included into a~singe superindex~``$i$'', so
that $\phi^i\equiv\phi^i(x)$. The repeated superindex implies
summation over its discrete part and integration over~$X$ with
respect to an appropriate measure. The partial derivatives with
respect to $\phi^i$ are understood as functional ones, that is,
$\partial_i=\delta/\delta \phi^i(x)$. The role of the space of
smooth functions $C^\infty(M)$ is played by the space of local
functionals of f\/ields; in so doing, two functionals are considered
to be equivalent if they dif\/fer by boundary terms. In particular,
the equality $S(\phi)=0$ implies that the local functional $S$ is
given by the integral of a~total divergence. Notice that unlike the
smooth functions on a f\/inite dimensional supermanifold, the local
functionals do not form a supercommutative algebra.

With the condensed notation, any system of f\/ield equations can be
written as
\begin{equation}\label{AT}
T_a(\phi)=0 .
\end{equation}
As we do not assume the f\/ield equations to come from the least
action principle, the discrete parts of the superindices $i$ and
$a$, labelling respectively the components of f\/ields and equations,
may run through completely dif\/ferent sets. For example, the
relativistic wave equations~(\ref{FEquations}) take their values in
the spin-tensor f\/ields of type $(s-1,1)$, while the corresponding
f\/ields constitute a spin-tensor of type~$(s,0)$. In this case, the
inf\/inite dimensional supermanifold  $M$ of spin-tensor f\/ields is
purely bosonic (i.e., just a manifold) for integer spin and purely
fermionic for half-integer ones. Continuing the geometric
interpretation above, one can regard the dif\/ferential operators~$T$'s as components of a~section~$T=\{T_a\}$ of some inf\/inite
dimensional supervector bundle $\mathcal{E}$ over the base
supermanifold $M$. In~\cite{KazLS}, it was proposed to call
$\mathcal{E}\rightarrow M$ the \textit{dynamics bundle}. Then the
solutions to the f\/ield equations~(\ref{AT}) are identif\/ied with the
zero locus $\Sigma\subset M$ of the section $T$, the \textit{shell}
in the physical terminology. The f\/ield equations~(\ref{AT}) are
called \textit{regular} if any local function of f\/ields that
vanishes on~$\Sigma$ is proportional to the local functions~$T_a$
and their derivatives with respect to the local coordinates on~$X$.
Below we assume that the equations~(\ref{AT}) are regular. The
functional derivative of~$T$'s gives the operator $J_{ia}=\partial_i
T_a$ def\/ining the so-called universal linearization of the equations~(\ref{AT}). It can also be viewed as an operator def\/ining a
homomorphism $J: TM\rightarrow \mathcal{E}$ from the tangent bundle
of the space of f\/ields $M$ to the dynamics bundle, if one regards~$\partial_i$ as a covariant derivative associated to a f\/lat
connection in~$\mathcal{E}$.

 A vector f\/ield $Z=Z^i\partial_i$ on $M$,
i.e., a section of the tangent bundle $TM$, is called a
\textit{symmetry} of the f\/ield equations (\ref{AT}) if
\[
J(Z)|_{\Sigma}=0\quad \Leftrightarrow\quad Z^iJ_{ia}=A_a^bT_b
\]
for some $A$'a. The symmetries form a subalgebra in the Lie
superalgebra of all vector f\/ields on~$M$.  Let $\mathcal{E}^\ast$
denote the vector bundle dual to the dynamics bundle $\mathcal{E}$.
A section $P=\{P^a\}$ of~$\mathcal{E}^\ast$ is called an
\textit{adjoint symmetry} if
\begin{equation}\label{AP}
J^\ast(P)|_\Sigma=0\quad \Leftrightarrow\quad J_{ia}P^a =B_i^bT_b
\end{equation}
for some $B$'s.  Among the adjoint symmetries one can extract those
originating from the identities. A section $Q$ of $\mathcal{E}^\ast$
is said to generate an identity for the equations (\ref{AT})  if
\begin{equation}\label{AQ}
T_aQ^a=0 .
\end{equation}
As we have explained above, the  last equality should be understood
in the sense that the local functional of f\/ields $T_aQ^a$ is given
by the integral of a total divergence $\mathrm{div}\, j$. By
def\/inition, the current $j$ is conserved when evaluated on solutions
to~(\ref{AT}), that is, $\mathrm{div}\, j|_{\Sigma}=0$. Neither
identities nor conserved currents are def\/ined by relation~(\ref{AQ})
uniquely. The equivalence classes of identities that correspond to
equivalence classes of nontrivial conserved currents are called
\textit{characteristics}. This gives a linear bijection between the
spaces of characteristics and nontrivial conserved currents.  Taking
the functional derivative of both the sides of equality~(\ref{AQ}),
one can see that each identity $Q$ satisf\/ies the adjoint symmetry
condition  (\ref{AP}).

Consider now a homomorphism  $V:\mathcal{E}^\ast\rightarrow TM$,
where  $V$ is assumed to be def\/ined, like the universal
linearization $J$, by some matrix dif\/ferential operator whose
coef\/f\/icients are local functions of f\/ields. By def\/inition, the
homomorphism $V$ is called a \textit{Lagrange anchor}, if the
following diagram of maps becomes supercommutative upon restriction
to ${\Sigma}$:
\begin{equation}\label{DIAG}
\xymatrix{TM \ar[r]^J &\mathcal{E}\\
{\mathcal{E}^\ast}\ar[u]^{V} \ar[r]^{J^\ast}  & {T^\ast
M}\ar[u]_{V^\ast}}
\end{equation}
For the regular equations (\ref{AT}), the last condition can be
written explicitly as
\begin{equation}\label{AVT}
V^i_a\partial_iT_b-(-1)^{\epsilon_a\epsilon_b}V^i_b\partial_i T_a
=C^d_{ab}T_d
\end{equation}
for some $C$'s. Here $\epsilon_a\in \mathbb{Z}_2$ denotes the
Grassmann parity of the local function of f\/ields $T_a$. In the
particular case of Lagrangian equations, $T_i=\partial_iS=0$, the
dynamics bundle coincides with the cotangent bundle $T^\ast M$ and~(\ref{AVT}) is satisf\/ied with the identity map $V=\mathrm{id}:
TM\rightarrow TM$ and all~$C$'s vanishing. This is known as the
\textit{canonical} Lagrange anchor for Lagrangian equations. Notice
that even in the Lagrangian situation the system~(\ref{AVT}) may
admit a lot of other (non-canonical) solutions. Like symmetries and
adjoint symmetries, all they form a vector superspace.

By def\/inition, the spaces of symmetries and adjoint symmetries are
identif\/ied with the on-shell kernels of the horizontal maps in~(\ref{DIAG}). Form the on-shell supercommutativity of the diagram it
then follows  that the vector f\/ield~$Z=V(P)$ is a symmetry of
(\ref{AT}) for any adjoint symmetry~$P$. Thus, each Lagrange anchor
$V$ gives rise to a homomorphism{\samepage
\begin{equation}\label{AV}
V: \  \mathrm{AdSym}(T)\rightarrow \mathrm{Sym}(T)
\end{equation}
acting from the space of adjoint symmetries to the space of
symmetries of the equations~(\ref{AT}).}

A Lagrange anchor $V$ is said to be \textit{ strongly
integrable}\footnote{In \cite{KLS2}, we used the term
\textit{integrable} instead of \textit{strongly integrable}. A more
relaxed notion of integrability was formulated in our recent paper~\cite{KLS3}. To distinguish between these two versions of
integrability, from now on we reserve the term \textit{integrable}
for the Lagrange anchors satisfying the relaxed integrability
condition in the sense of~\cite{KLS3}.} if the following two
conditions are satisf\/ied
\begin{equation}\label{AAn}
[V_a,V_b]=C_{ab}^dV_d ,\qquad
(-1)^{\epsilon_a\epsilon_c}\big(C_{ab}^dC^e_{cd}+V_c^i\partial_iC_{ab}^e\big)+{\rm cycle}(a,b,c)=0 .
\end{equation}
The f\/irst equation means that  $\mathrm{Im} V \subset TM$  is an
integrable distribution on~$M$. In case $\mathrm{Ker}\, V=0$, the
second condition in~(\ref{AAn}) follows from the f\/irst one by the
Jacobi identity for the supercommutator of the local vector f\/ields
$V_a=V_a^i\partial_i$. The integrability conditions~(\ref{AAn})
have a nice geometric interpretation as def\/ining a \textit{Lie
superalgebroid} over $M$. For a general discussions of Lie
algebroids we refer the reader to~\cite{K}. Upon this interpretation
the Lagrange anchor is identif\/ied with the anchor of a Lie
superalgebroid $V: \mathcal{E}^\ast\rightarrow TM$ and the Lie
superalgebra structure on the sections of $\mathcal{E}^\ast$ is
def\/ined by the bracket
\[
[e_a, e_b]=C_{ab}^d e_d ,
\]
where $\{e_a\}$ are frame sections in $\mathcal{E}^\ast$. The
def\/ining relation (\ref{AVT}) for the Lagrange anchor can then be
reinterpreted as the closedness condition for the
$1$-$\mathcal{E}$-form $T$ with respect to the Lie algebroid
dif\/ferential, $d_{\mathcal{E}}T=0$; in so doing, the integrability
condition (\ref{AAn}) is expressed by the operator equality
$d^2_\mathcal{E}=0$. It is not hard to see~\cite{KLS2} that for a
strongly integrable Lagrange anchor the generators of identities
(\ref{AQ}) form a subalgebra in the full Lie algebra of sections of
$\mathcal{E}^\ast$. The canonical Lagrange anchor of Lagrangian
equations gives an example of strongly integrable Lagrange anchor.
It corresponds to the tangent Lie algebroid $\mathrm{id}:
TM\rightarrow TM$. Although the theory of Lie algebroids and
groupoids is a fascinating area of modern dif\/ferential geometry, it
by no means covers or substitutes  the concept of a Lagrange anchor
completely, as the strong integrability condition (\ref{AAn}) is too
restrictive and is not actually needed in many f\/ield-theoretical
applications.

\section[Lagrange anchor for the Bargmann-Wigner equations]{Lagrange anchor for the Bargmann--Wigner equations}\label{appendixB}

Unfolding the condensed notation of Appendix~\ref{appendixA}, one can see that the
universal linearization of the f\/ield equations (\ref{FEquations})
and the Lagrange anchor (\ref{V}) are def\/ined by the following
operators:
\begin{gather}
J^{\beta\beta_1\ldots\beta_{2s-1},}{}^{\dot\alpha}_{\phantom{\dot\alpha}\alpha_1\ldots\alpha_{2s-1}}(x,z)=\delta_{\alpha_1}^{(\beta_1}\cdots
\delta_{\alpha_{2s-1}}^{\beta_{2s-1}}(\partial^z)^{\dot\alpha\beta)}\delta(x-z) ,\nonumber\\
V^{\gamma}{}_{\dot{\gamma}_1\ldots\dot{\gamma}_{2s-1},\beta\beta_1\ldots\beta_{2s-1}}(z,y)=
(-i)^{2s}\delta_{(\beta}^{\gamma}(\partial^z)_{\beta_1\dot\gamma_1}\cdots(\partial^z)_{\beta_{2s-1})\gamma_{2s-1}}\delta(z-y) .\label{ULA}
\end{gather}
There are also operators corresponding to the complex conjugate
equations (\ref{FEquations}). Their integral kernels are obtained by
complex conjugation of (\ref{ULA}).

Since both the operators in (\ref{ULA}) are independent of f\/ields,
the def\/ining property for the Lagrange anchor is fulf\/illed if\/f the
left hand side of (\ref{AVT}) is equal to zero identically, that is,
all~$C$'s must vanish.  This amounts to the equality
\begin{gather}
\int \! d^4z
J^{\beta\beta_1\ldots\beta_{2s-1},\dot\alpha}{}_{\alpha_1\ldots\alpha_{2s-1}}(x,z)
V^{\gamma}{}_{\dot{\gamma}_1\ldots\dot{\gamma}_{2s-1},\beta\beta_1\ldots\beta_{2s-1}}(z,y)-(i)^{4s}\overline{\big(\alpha\!\leftrightarrow\!\gamma
,x\!\leftrightarrow\! y\big)}=0 .\!\!\!\label{LA2}
\end{gather}
On substituting (\ref{ULA}) into (\ref{LA2}), we get
\begin{gather}
\Big\{(\partial^y)^{\gamma\dot\alpha}(\partial^{y})_{\alpha_1(\dot\gamma_1}
 \cdots(\partial^{y})_{\alpha_{2s-1}\dot\gamma_{2s-1})}\delta(x-y)-(-1)^{2s}\big(x\leftrightarrow
y\big)\Big\} \nonumber\\
 {} +\Bigg[\sum_{p=1}^{2s-1}
\delta^{\gamma}_{\alpha_p}(\partial^y)^{\beta\dot\alpha}(\partial^y)_{\beta\dot\gamma_p}
(\partial^{y})_{\alpha_1(\dot\gamma_1}\cdots(\partial^{y})_{\alpha_{p-1}\dot\gamma_{p-1}}(\partial^{y})_{\alpha_{p+1}\dot\gamma_{p+1}}
\cdots(\partial^{y})_{\alpha_{2s-1}\dot\gamma_{2s-1})}\delta(x-y) \nonumber\\
{} -(-1)^{2s}\big(\alpha\leftrightarrow\dot\gamma,
\dot\alpha\leftrightarrow\gamma, x\leftrightarrow
y\big)\Bigg]=0 . \label{PP}
\end{gather}
Due to the symmetry properties of the derivatives of Dirac's
$\delta$-function, the terms in the braces cancel each other.
Transferring now all the partial derivatives in the square brackets
to $y$'s  and using the identities
\[
(\partial^y)^{\beta\dot\alpha}(\partial^y)_{\beta\dot\gamma_p}=
\frac{1}{2}\delta^{\dot\alpha}_{\dot\gamma_p}(\partial^y)^{\beta\dot\beta}(\partial^y)_{\beta\dot\beta} ,
\qquad
(\partial^y)^{\gamma\dot\beta}(\partial^y)_{\alpha_p\dot\beta}=
\frac{1}{2}\delta_{\alpha_p}^{\gamma}(\partial^y)^{\beta\dot\beta}(\partial^y)_{\beta\dot\beta} ,\]
one can bring the left hand side of (\ref{PP}) to the form
\begin{gather*}
\frac12\Bigg\{\sum_{p=1}^{2s-1}
\Big(\delta^{\gamma}_{\alpha_p}\delta^{\dot\alpha}_{(\dot\gamma_p}-\delta^{\gamma}_{\alpha_p}
\delta^{\dot\alpha}_{(\dot\gamma_p}\Big)(\partial^y)^{\beta\dot\beta}(\partial^y)_{\beta\dot\beta}  \\
\qquad{} \times(\partial^{y})_{\alpha_1\dot\gamma_1}\cdots(\partial^{y})_{\alpha_{p-1}\dot\gamma_{p-1}}(\partial^{y})_{\alpha_{p+1}\dot\gamma_{p+1}}
\cdots(\partial^{y})_{\alpha_{2s-1}\dot\gamma_{2s-1})}\delta(x-y)\Big\} .
\end{gather*}
The last expression is equal to zero identically, and verif\/ication
of the def\/ining property (\ref{AVT}) for the Lagrange anchor
(\ref{V}) is completed.

\subsection*{Acknowledgments}

We are thankful  to G.~Barnich, and E.~Skvortsov for illuminating
discussions on various issues related to the subject of this paper
and for relevant references. We also benef\/ited from the valuable
comments by S.~Anco and three anonymous referees that helped us to
improve the initial version of the manuscript. The work was
initiated during our visit to the Erwin Shr\"odinger Institute
(Vienna, October--November 2010) and we appreciate its hospitality.
The work was partially supported by the RFBR grant No 09-02-00723-a
and also by Russian Federal Agency of Education under the State
Contract no P789. AAS appreciates the f\/inancial support from Dynasty
Foundation, SLL acknowledges partial support from the RFBR grant
11-01-00830-a.

\pdfbookmark[1]{References}{ref}
\LastPageEnding

\end{document}